\documentclass[superscriptaddress,amsmath,amssymb,nofootinbib,twocolumn]{revtex4}

\usepackage{setspace}
\usepackage{graphicx}  
\usepackage{dcolumn}   
\usepackage{bm}        
\usepackage{rotating}
\usepackage{hyperref}
\usepackage{xcolor}
\usepackage{adjustbox}

\newcommand{\sectionname}[1]{ \noindent{\bfseries #1}.---}

\begin{document}




\title{
Dissipative Effects as New Observables for Cosmological Phase Transitions 
}
\author{Huai-Ke Guo}
\email{guohuaike@ucas.ac.cn}
\affiliation{
International Centre for Theoretical Physics Asia-Pacific, 
University of Chinese Academy of Sciences, 100190 Beijing, China
} 
\date{\today}


\begin{abstract}
We show that dissipative effects during cosmological first order phase transitions
lead to a frequency-dependent suppression for the usually 
dominant gravitational wave production from sound waves, through an analytical modelling of the
source based on the sound shell model. This damping effect is more pronounced 
for high frequencies or small scales, and modifies the spectral shape and possibly the
peak frequency. These modifications can be used to reveal more information about 
the underlying particle interactions, serving as a way of breaking the parameter degeneracy 
that plagues particle physics studies based on the perfect fluid approximation.
\end{abstract}


\maketitle

\sectionname{Introduction}
The direct detection of gravitational waves from compact object coalescences 
by the LIGO and Virgo collaboration~\cite{LIGOScientific:2016aoc} has revived the interest
in the searches and theoretical studies of a stochastic background of gravitational waves of cosmological origin (see, e.g., \cite{Cai:2017cbj,Caprini:2018mtu,Christensen:2018iqi,Caldwell:2022qsj} for reviews),
which if discovered would be another milestone and can play a role for fundamental physics~\cite{Bian:2021ini,Caldwell:2022qsj,LISACosmologyWorkingGroup:2022jok} 
similar to what the cosmic microwave background radiation means for modern cosmology. One important class of this kind
is that from cosmological first order phase transitions in the early universe~\cite{Caprini:2015zlo,Caprini:2019egz,Mazumdar:2018dfl,Hindmarsh:2020hop,Athron:2023xlk},
which is directly connected to the underlying particle physics and thus to potentially new physics. 
Gravitational waves from phase transitions have already been
searched for at LIGO~\cite{Romero:2021kby} corresponding to new physics at the scale of 
$\mathcal{O}(10^3-10^6) \text{TeV}$~\cite{Badger:2022nwo} well beyond what current high energy colliders can reach, 
and at pulsar timing array experiments~\cite{NANOGrav:2021flc,Xue:2021gyq} corresponding to the QCD scale. 
More recently, the pulsar timing array experiments have reported evidence for a stochastic background of gravitational waves~\cite{Xu:2023wog,NANOGrav:2023gor,Reardon:2023gzh,EPTA:2023fyk}, with phase transition gravitational waves being one of the potential sources~\cite{NANOGrav:2023hvm,EPTA:2023xxk}.
At the electroweak scale, where current colliders operate, phase transition gravitational
waves plays a complementary role to direct searches of new particles at collider (see, e.g.,~\cite{Caldwell:2022qsj} for a review) 
and can help to pin down the origin of the electroweak symmetry breaking as well as 
the origin of the baryon asymmetry in the universe~\cite{Morrissey:2012db}, thus making it an important goal for  
future space-based gravitational wave detectors LISA~\cite{LISA:2017pwj,Caprini:2015zlo}, Taiji~\cite{Hu:2017mde,Ruan:2018tsw,Taiji-1} and 
Tianqin~\cite{TianQin:2015yph,Luo:2020bls,TianQin:2020hid}. The cosmological first order phase transitions 
can also provide new formation mechanisms for primordial black hole dark matter~\cite{Baker:2021nyl,Kawana:2021tde,Liu:2021svg,Huang:2022him,Xie:2023cwi},
generate new curvature perturbations~\cite{Liu:2022lvz}, and
produce primordial magnetic field that can potentially explain the observed magnetic field in the voids~\cite{Di:2020kbw,Yang:2021uid}, among others.

The above goals hinge on a precise understanding of the dynamics and kinematics of the phase transition, and
significant progress has been witnessed in recent years.
It is now generally accepted that there are mainly three mechanisms for gravitational wave
production during such a transition: bubble collisions~\cite{Kosowsky:1992rz,Kosowsky:1992vn,Jinno:2016vai}, 
sound waves~\cite{Hindmarsh:2013xza,Hindmarsh:2015qta} and
magnetohydrodynamic (MHD) turbulence~\cite{Caprini:2015zlo,Kahniashvili:2008pf,Kahniashvili:2008pe,Kahniashvili:2009mf,Caprini:2009yp,Kisslinger:2015hua,Pol:2019yex}. 
For transitions in a thermal plasma, the acoustic production from sound waves is generally believed to be the dominant, 
while that from the MHD turbulence is subdominant and quite 
uncertain as of now. This makes a precise prediction of the acoustic production of gravitational waves
especially important. Previous studies, however, rely on the perfect fluid approximation to the plasma, with
dissipative effects largely neglected, though of course they are naturally included in studies of turbulence (see, e.g.,~\cite{Caprini:2009yp,Dahl:2021wyk}).
This results in, among others, the problem of parameter degeneracy, that 
the spectrum depends only on a set of bulk fluid parameters, which in turn translates into the dependence on 
the set of phase transition parameters: the dimensionless energy release normalized by the radiation energy density $\alpha$,
the typical inverse time duration $\beta$ or the mean bubble separation $R_{\ast}$, 
the wall velocity $v_w$ and the transition temperature $T_{\ast}$, 
while many particle physics models can lead to 
the same parameter values. Thus it is highly desirable to find ways that can break the parameter degeneracy.

In this work, we show that dissipative effects, which have largely been neglected in previous studies,
can serve as one way of breaking the degeneracy, and makes possible the probing of very weak particle interactions,
in cases where the effect of dissipation is strong.


\sectionname{Dissipative effects in an imperfect fluid} 
For the generally dominant gravitational wave production from sound waves,
when the scalar field driving the phase transition no longer plays a significant
role and where possible electromagnetic fields are neglected, the matter content 
in the universe consists generally of a plasma of relativistic particles,
possible non-relativistic particles, or others. The energy momentum tensor of 
such a system is described in previous studies by the well known
perfect fluid form $p g^{\mu\nu} + (\rho + p) U^{\mu} U^{\nu}$, where $U^{\mu}$ is the
velocity four-vector of the fluid, and $p$ and $\rho$ are the pressure and energy density 
both measured in a locally comoving Lorentz frame (i.e., ${\bf U}=0$)  at a certain instant of time. However, in the presence 
of dissipative effects, which drive the system to a new equilibrium state, 
the energy momentum tensor needs to be modified by including a new term $\Delta T^{\mu\nu}$~\cite{Weinberg:1971mx}, which
are described by three positive parameters: the shear viscosity $\mu$, bulk viscosity $\zeta$ and thermal 
conduction $\chi$. Dissipations of these kinds are well known in Newtonian fluid mechanics and also in
relativistic hydrodynamics (see e.g.,~\cite{landau1987fluid}).
In a Lorentz frame comoving with the fluid at a spatial point $x^i$ and at a specific time $x^0=t$, 
it takes the following form~\cite{Weinberg:1971mx}
\begin{eqnarray}
&& \Delta T^{ij} = - \mu \left(\frac{\partial U_i}{\partial x^j} + \frac{\partial U_j}{\partial x^i} - \frac{2}{3} \delta_{ij} \triangledown \cdot {\bf U}\right) 
                     - \zeta\ \delta_{ij} \triangledown \cdot {\bf U} , \nonumber \\
&& \Delta T^{i0} = - \chi \left(\frac{\partial T}{\partial x^i} + T \dot{U}_i \right) .
\end{eqnarray}
Going back to a generic frame, the equations driving the evolution of the system, more specifically of $p$, $\rho$, ${\bf v}$, etc, can 
then be obtained by the conservation of energy and momentum, and also by the conservation of conserved quantum numbers
present in the system. 
The part of the energy momentum tensor that generates gravitational waves is $a^2 (p + \rho) \gamma^2 v^i v^j$, 
where $v^i \equiv d x^i/d\eta$ with $\eta$ the comoving time. To facilitate an analytical insight into the 
underlying physics, we assume in the following that the perturbations 
caused by the phase transition are small such that in calculating the energy momentum tensor we neglect the
fluctuations of $p$ and $\rho$. So the key task is on the calculation of the stochastic velocity field ${\bf v}(\eta, {\bf x})$.
In the absence of dissipation, the sound equation leads to the following solution for the velocity field in an expanding universe~\cite{Guo:2020grp}
\begin{eqnarray}
v^i(\eta, \mathbf{x}) = \int \frac{d^3 q}{(2\pi)^3} 
\left[
  {v}^i_{\mathbf{q}} e^{- i \omega \eta + i \mathbf{q} \cdot \mathbf{x}} 
+
c.c.
\right] ,
\end{eqnarray}
where $\omega = q c_s$, with ${\bf q}$ the comoving wavenumber and $c_s$ the speed of sound, which is the dispersion relation in the absence of dissipation. 
The presence of dissipation has the effect of converting kinetic energy of the fluid into heat, leading then to a damping of sound waves, thus making
${v}^i_{\mathbf{q}}$ dependent on $\eta$.
For the longitudinal sound waves excited by the expanding bubbles with wavenumber ${\bf q}$, the 
amplitude of the velocity Fourier component is damped exponentially in the following way~\cite{Weinberg:1971mx}
\begin{eqnarray}
  v^i_{\mathbf{q}}(\eta) \propto \text{exp}\left[- \int \Gamma(\mu, \zeta, \xi) d \eta\right] .
\end{eqnarray}
The detailed form of $\Gamma$ was derived in~\cite{Weinberg:1971mx}, with the key property that $\Gamma \propto q^2$, 
a result well known from Newtonian fluid mechanics (see e.g.,~\cite{landau1987fluid}). Thus perturbations of smaller scales or larger frequencies are
more damped by the presence of dissipation.

\sectionname{Velocity field and power spectrum} 
With the equations of motion setup for the system and the effect of dissipation included, we need the initial conditions
to obtain the velocity field.
Here the non-zero velocity field is excited by the interaction between the plasma and expanding bubbles: as each bubble
expands, the plasma surrounding it is stirred.  Because there
are many bubbles in a Hubble volume, a precise determination of the velocity field would be by solving the 
fluid equations together with those governing the expansion and destructions of the bubbles, i.e., the evolution of the scalar field responsible
for the transition. 
However, in the case where the velocity is small one can add linearly the
contributions from all bubbles, which is the essence of the sound shell model~\cite{Hindmarsh:2016lnk, Hindmarsh:2019phv} 
(see~\cite{Guo:2020grp,Cai:2023guc,RoperPol:2023dzg} for generalizations),
such that an analytical determination of velocity power spectrum can be achieved.
Then in the sound shell model, the coefficient of each Fourier component ${v}^i_{\mathbf{q}}(\eta)$ is
obtained by linearly superposing the perturbations from a total of, say $N_b$, bubbles ever nucleated and destroyed
upon collision with another bubble
\begin{eqnarray}
  v_{\bf{q}}^i(\eta) = \sum_{n=1}^{N_b} v_{\bf{q}}^{i (n)} 
  \text{exp}\left[- \int_{\eta_d^{(n)}}^{\eta} \Gamma d \bar{\eta}\right] \theta(\eta - \eta^{(n)}_d) ,
\end{eqnarray}
where $\eta^{(n)}_d$ denotes the destruction time of the n'th bubble at which the stirred velocity
starts contributing to sound waves, neglecting for simplicity possibly forced motion of the sound shells~\cite{Cai:2023guc}. 
The $\eta$-independent $v_{\bf{q}}^{i (n)}$ can be calculated by solving the velocity profile~\cite{Espinosa:2010hh} for 
the n'th one that is nucleated at time $\eta_s^{(n)}$ and at location ${\bf x}^{(n)}$~\cite{Guo:2020grp} giving then
$v_{\bf{q}}^{j (n)} = i \hat{q}^j \left(\eta_{lt}^{(n)}\right)^3 
e^{i \omega \eta^{(n)}_d - i \mathbf{q} \cdot \mathbf{x}^{(n)}}A\left(q \eta_{lt}\right)$,
where $\eta_{lt}^{(n)} = \eta^{(n)}_d - \eta^{(n)}_s$, which is the conformal lifetime of the n'th bubble, and
$A$ is a function with an absolute value that peaks at $q \eta_{lt} \sim \mathcal{O}(1)$.
The physical meaning of this equation is quite clear: at time $\eta_d^{(n)}$ when the $n$'th bubble is 
destroyed, the initial velocity perturbation is matched onto freely propagating sound waves, giving then its contribution
to the Fourier component with an amplitude that is damped over the following time due to dissipation. Since different 
bubbles contribute at different times, the corresponding component of sound waves get damped
at different times accordingly.
\begin{figure}
\centering
\includegraphics[width=0.48\columnwidth]{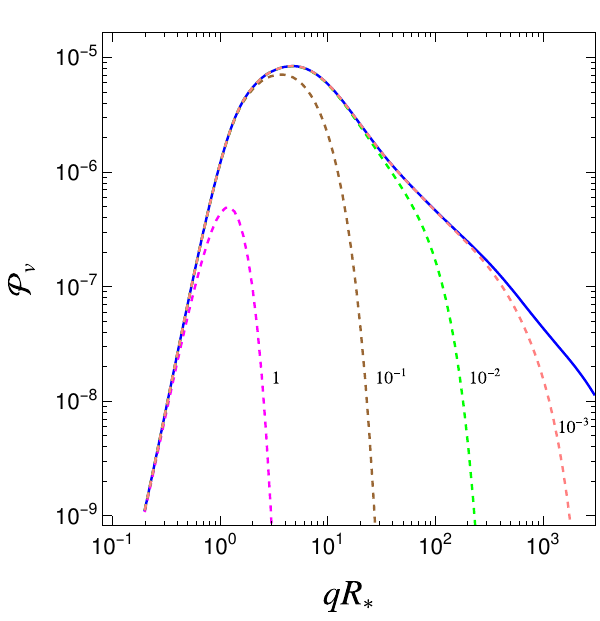}
\includegraphics[width=0.48\columnwidth]{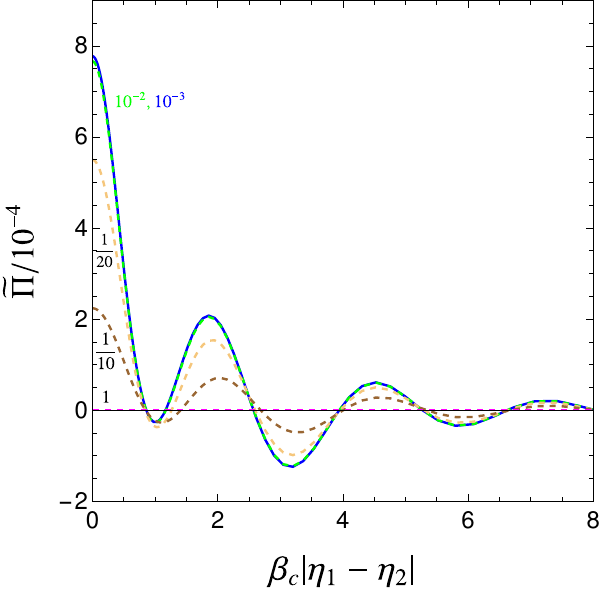}
\caption{
\label{fig:aux}
Left: velocity power spectrum as a function of the dimensionless wavenumber $q R_{\ast}$.
Right: source auto-correlation as a function of the dimensionless conformal time difference $\beta_c |\eta_1 - \eta_2|$ 
for $q R_{\ast} = 10$. The numbers in the plots denote the ratio $d_D/R_{\ast}$.
In both plots, we fix $\alpha=0.0046$, $v_w = 0.92$ as an example.
}
\end{figure}

The summation of $N_b$, bubbles leads to a velocity $v^i_{\bf q}(\eta)$ that has a stochastic nature:
the bubbles are nucleated and destroyed at different random times and at different random locations, resulting thus in
a random velocity field. This stochastic nature, which is also classical, is similar to that encountered in the 
standard cosmological perturbation theory where however
the randomness originates from the quantum fluctuations of the inflaton (see, e.g.,~\cite{Weinberg:2008zzc}).
For such stochastic fields, meaningful quantities are their averages. Since the number of bubbles $N_b$ is generally
large, and according to the central limit theorem, the velocity is Gaussian to a good approximation, in which
case the fundamental average is the two-point correlator. With the notation of Fourier transform
$\tilde{v}_{\mathbf{q}}^i(\eta) = \int d^3 {\bf x}\ e^{-i {\bf q \cdot x}} v^i(\eta, {\bf x})$.
The fundamental two-point correlator takes a form that is a generalized version of that in the absence of dissipations~\cite{Hindmarsh:2019phv,Guo:2020grp}
\begin{eqnarray}
&& \langle \tilde{v}_{\mathbf{q}}^i(\eta_1) \tilde{v}_{\mathbf{k}}^{j \ast}(\eta_2)\rangle 
  = 2 \pi^2 q^{-3} \delta^{3}(\mathbf{q}-\mathbf{k}) \hat{q}^i \hat{k}^j  \nonumber \\
  && \hspace{2.5cm} \times \mathcal{P}_v(q, \eta_1, \eta_2) \cos[\omega (\eta_1 - \eta_2)] ,
\end{eqnarray}
where the delta function comes from the overall homogeneity of the universe and the velocity power spectrum 
$\mathcal{P}_v$ can be shown to be 
\begin{eqnarray}
&& \mathcal{P}_v(q, \eta_1, \eta_2) 
= \frac{q^3}{\pi^2} \int d \eta_{\text{lt}} \int d \eta_d \left[ P(\eta_{\text{lt}}, \eta_d) \frac{N_b}{V} \right] 
\nonumber \\
  && \hspace{0.8cm} \times \eta_{\text{lt}}^6 |A(q \eta_{\text{lt}})|^2 
  \text{exp}\left[- \int_{\eta_d}^{\eta_1} \Gamma d t - \int_{\eta_d}^{\eta_2} \Gamma d t\right] .
\end{eqnarray}
Here $P(\eta_{\text{lt}}, \eta_d)$ is the probability density function of bubbles with lifetime $\eta_{\text{lt}}$ and destruction time $\eta_d$.
Marginalizing over $\eta_d$ gives the lifetime distribution $P(\eta_{\text{lt}})$ that can be derived analytically~\cite{Hindmarsh:2019phv,Guo:2020grp} 
or numerically from a simulation of bubble nucleation~\cite{Hijazi:2022uzc,Guo:2023gwv}, while the full distribution $P(\eta_{\text{lt}}, \eta_d)$ 
can be obtained straightforwardly from results of these numerical simulations if not possible analytically.
In the absence of dissipations, $\mathcal{P}_v$ is independent of $\eta_1$ and $\eta_2$ and the right hand side of the velocity correlator 
depends on $\eta_1$ and $\eta_2$ through only the combination $(\eta_1 - \eta_2)$, meaning that the velocity field is stationary.
Inclusion of dissipations, however, makes $\mathcal{P}_v$ dependent on $\eta_1$ and $\eta_2$ separately, and thus renders the velocity
field non-stationary. Physically the presence of this nonstationarity is apparent 
as the damping caused by dissipation accumulates over time and thus depends on the absolute time when each velocity perturbation kicks in. 

In light of the $q^2$ dependence of the decay rate $\Gamma$, it is convenient to define an effective damping length 
$\int_{\eta_d}^{\eta_1} \Gamma d t = q^2 d^2_D(\eta_d, \eta_1)$ and then the exponent becomes 
$q^2 [d^2_D(\eta_d, \eta_1) + d^2_D(\eta_d, \eta_2)] \equiv q^2 d^2_D(\eta_d, \eta_1, \eta_2)$. 
In cases where the bubbles all disappeared within a short time range, we can neglect the variation among the bubble destruction times, and simply
use a typical time $\eta_{\ast}$. This leads to a result similar to that obtained without dissipations $\mathcal{P}_v(q)$,
\begin{eqnarray}
\mathcal{P}_v(q, \eta_1, \eta_2) =  \text{exp}\left[- q^2 d^2_D(\eta_{\ast}, \eta_1, \eta_2) \right] \mathcal{P}_v(q) .
\end{eqnarray}
The effect of dissipations on the velocity power spectrum are shown in the left panel of Fig.~\ref{fig:aux}, for the simplified case of a constant 
$d_D(\eta_{\ast}, \eta_1, \eta_2)$ in unit of the typical scale of the phase transition, the mean bubble separation $R_{\ast}$, for illustration. 
One can clearly see the damping of the high frequency tail for $d_D/R_{\ast} = 10^{-3}$, and for larger values of $d_D$ the damping 
effect extends more broadly to lower frequencies, and can even shift the peak to lower values.


\sectionname{Damping of gravitational waves}
The power spectrum of the stochastic gravitational waves generated by the stochastic velocity field 
can be obtained by firstly solving the gravitational wave amplitude in terms of the source and then 
calculating its correlator which is reduced to sums of product of the velocity correlators, and 
eventually becomes~\cite{Guo:2020grp}:
\begin{eqnarray}
\mathcal{P}_{\text{GW}}(\eta, k)
&=& \frac{32 G^2 [\left( \bar{\rho} + \bar{p} \right) \bar{U}_f^2]^2}{3 a^2 H^2} (k R_{\ast})^3 
\int_{\tilde{y}_s}^{\tilde{y}} d \tilde{y}_1 \int_{\tilde{y}_s}^{\tilde{y}} d \tilde{y}_2 
  \nonumber \\
&& \times 
\left(\frac{\partial \tilde{y}}{\partial \tilde{\eta}}\right)^2
\frac{\partial G(\tilde{y}, \tilde{y}_1)}{\partial \tilde{y}}
\frac{\partial G(\tilde{y}, \tilde{y}_2)}{\partial \tilde{y}} 
\frac{a(\eta_s)^8}{a^2(\eta_1) a^2(\eta_2)}  \nonumber \\
&& \times \frac{\tilde{\Pi}^2(k R_{\ast}, k \eta_1, k \eta_2)}{k^2} ,
\end{eqnarray}
where $G(\tilde{y}, \tilde{y}_{1/2})$ is the Green's function; $\tilde{y}=f(k) a/a(\eta_s)$
with $f(q)$ a factor depending on the expansion rate of the universe; 
$\bar{U}_f$ is the root-mean-square fluid velocity. The key in this equation is the dimensionless source auto-correlator
\begin{eqnarray}
&&\tilde{\Pi}^2
=
\frac{\pi}{2} \frac{1}{\bar{U}_f^4} 
\int d^3 \tilde{q} \mathcal{P}_v(\tilde{q}) \mathcal{P}_v(\tilde{\bar{q}}) \frac{(1-\mu^2)^2}{\tilde{q} \tilde{\bar{q}}^5}
e^{- (q^2 + \bar{q}^2) d_D^2} \nonumber \\
  && \hspace{0.8cm} \times
\cos\left[c_s \tilde{q} \frac{\beta_c (\eta_1 - \eta_2)}{\beta_c R_{\ast}} \right]
\cos\left[c_s \tilde{\bar{q}} \frac{\beta_c (\eta_1 - \eta_2)}{\beta_c R_{\ast}} \right] .
\label{eq:PiExplicit}
\end{eqnarray}
where $\bar{q}=|\mathbf{q}-\mathbf{k}|$, $\mu = \hat{\mathbf{q}} \cdot \hat{\mathbf{k}}$, $\tilde{q} = q R_{\ast}$ and
$\beta_c$ is the comoving version of $\beta$~\cite{Guo:2020grp}.
The integral in $\mathcal{P}_{\text{GW}}$ can equivalently be transformed into one over $(\eta_1 - \eta_2)$ and 
another linear combination~\cite{Guo:2020grp}, which is more useful for a stationary source.
The nonstationarity in the velocity power spectrum propagates into $\tilde{\Pi}^2$, making it also non-stationary and
also making the integral more complicated. Despite this, qualitative physical insights can be gained:
the highly oscillatory trigonometric functions force $\eta_1$ to be close to $\eta_2$, the same as
that in the absence of the smoothly varying dissipation term, and the exponential factor leads to a reduced 
auto-correlator. In the right panel of Fig.~\ref{fig:aux}, we show the source auto-correlator 
for several constant values of the ratio $d_D/R_{\ast}$. We can again see the reduced overall amplitude of the 
auto-correlator due to dissipation. Note that the assumption of a constant $d_D$ in the above example results in a 
stationary source correlator, while generically the correlator depends on $\eta_1$ and $\eta_2$ in a more complicated way.

\begin{figure}[t]
\centering
\includegraphics[width=0.9\columnwidth]{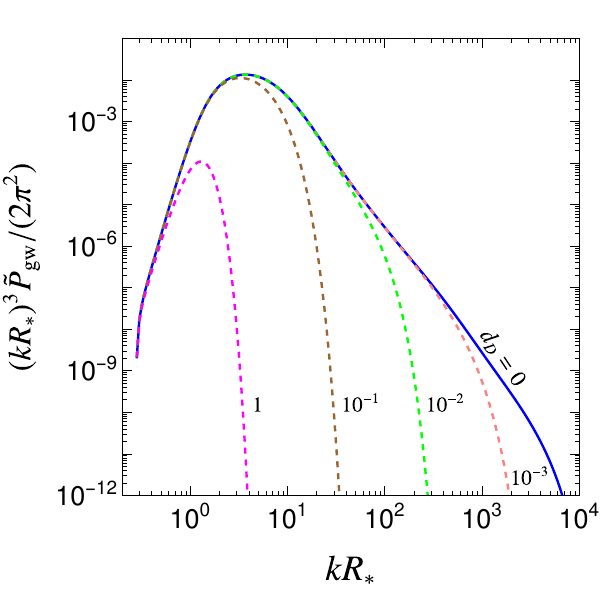}
\caption{
\label{fig:Pgw}
The dimensionless gravitational wave spectrum before redshift 
as a function of the dimensionless wavenumber $k R_{\ast}$, for several constant values of $d_D/R_{\ast}$.
}
\end{figure}

With above dampings found for the velocity power spectrum and the source auto-correlator, it is no wonder
that similar dampings show up in the eventual gravitational wave spectrum, and this is shown in Fig.~\ref{fig:Pgw}
for the same parameter choices as in previous plots for $d_D/R_{\ast} = 0, 10^{-2}, 10^{-1}$ and $1$. Here the quantity 
plotted captures all the $k$-dependence of $\mathcal{P}_{\text{GW}}(\eta, k)$ whereas unrelated overall factors are neglected. 
The message carried by this plot is the central result of this work: dissipations lead to a sharp suppression of the 
gravitational wave spectrum, usually acting at the high frequency tail and possibly extending to the lower frequency regime. 
While dampings at the high frequency tail affect the UV spectral shape,
those at low frequencies can even shift the peak frequency to smaller values.
Exactly at what value of the frequency does the damping start to appear and how much the spectrum is damped depend highly on the values
of $\mu, \zeta$ and $\chi$ and their possible time variations as the universe expands. 
For the simplified example where $d_D$ are constants, the source becomes stationary and 
the accumulation of gravitational wave power over time is mostly decoupled from the integral over the
short auto-correlation time $(\eta_1 - \eta_2)$, which results then in the factorized form of the spectrum, i.e.,
$\mathcal{P}_{\text{GW}} \propto (k R_{\ast})^3 \tilde{P}_{\text{gw}} \Upsilon(\tau_{\text{sw}})$~\cite{Guo:2020grp} 
with $\tau_{\text{sw}}$ the lifetime of sound waves. For more realistic scenarios where $d_D$ is a function of both 
$\eta_1$ and $\eta_2$, the situation is more complicated, as the spectrum generated at a later time has a different
shape as that from an earlier time, and above factorized form of the spectrum does not generally exist.

\sectionname{Lifetime of sound waves and time scales}
The interplay between the suppressed production of gravitational waves due to the expansion of the universe, as 
captured by the function $\Upsilon$, and the damping due to dissipation in realistic cases where $d_D$ varies with
time leaves interesting imprints on the shape as well as on the amplitude of the gravitational wave spectrum. 
The damped spectral shape for the affected frequency range varies over time and thus the eventual shape is a combined
result in which the details of the dissipation are recorded.
In addition, for cases where the effective damping length $d_D$ is large, the damping, which now reduces the amplitude
across a larger frequency range, together with the dilution effect can strive to determine a new effective lifetime of 
the sound waves, beyond which negligible gravitational waves are produced. This is in contrast to the widely adopted
lifetime of sound waves, $R_{\ast}/\bar{U}_f$~\cite{Pen:2015qta,Hindmarsh:2017gnf}, which corresponds to the onset of MHD turbulence. 
More interestingly, in cases where dissipation is strong and thus the Reynolds number is significantly reduced, the turbulence 
might never have been generated, and it is solely dissipation that determines the lifetime of sound waves.

From the above analysis, 
one can identify the following time scales without performing a dedicated numerical integration to get an 
analytical insight: (1) auto-correlation time, which is typically very short and thus decoupled from the others;
(2) mean free time of the particles leading to the dissipation, which and whose time evolution
determine eventually the value and the time variation of the effective damping length $d_D$; 
(3) the onset time of the MHD turbulence, which marks the end of the acoustic production of gravitational waves;
(4) the Hubble time, or the expansion rate of the universe when the sound waves are active. 
The final gravitational wave spectrum from sound waves is thus a complicated product as a result of the
intertwining of these time scales and the underlying effects, which is model dependent and also depends
on the cosmological context. Seemingly like messy complications, this is really a
blessing as it breaks the parameter degeneracy that plagues the study based on the perfect fluid approximation.


\sectionname{Experimental detection and microscopic origin} 
The modified spectral shape and amplitude, as parametrized by $\mu$, $\zeta$ and $\chi$, can be readily searched for at gravitational wave detectors,
to extend previous searches at LIGO, Virgo and KAGRA~\cite{Romero:2021kby}, and at PTA experiments~\cite{NANOGrav:2023hvm}.
For future space-based interferometers such as LISA, Taiji and Tianqin, which target
the most important electroweak phase transition, more correlation and complementarity between experimental detection~\cite{Gowling:2021gcy,Gowling:2022pzb} 
and traditional particle physics studies, such as direct searches at the high energy colliders, can be readily explored.
Measurements of this kind open new portals into studies of the underlying microscopic particle interactions. 
The particle information gained by the measurement of $\mu$, $\zeta$ and $\chi$ differs from the studies based on the perfect fluid approximation 
in that the former is caused by particles with long mean free path in the plasma, thus with very weak 
interactions, while the latter is from the bulk fluid parameters in the perfect fluid approximation. This selects out 
potentially important applications to the probing of dark matter and other more weakly interacting particles which are abundant in 
physics beyond the standard model.
Thus studies of the particle interactions from the dissipative observables are 
complementary to their traditional direct studies which are effective when the interactions are strong.

Aside from gravitational waves, also known as tensor mode perturbation in the cosmological perturbation theory,
dissipations also suppress scalar perturbations, such as the density perturbations. Indeed
the same kind of damping is known in the cosmological perturbation theory as Silk damping~\cite{Silk:1967aha} which
suppresses the power spectrum of the CMB temperature anisotropy at large multipole moment $l$, due to the increasing
diffusion length of the photons at recombination. 
Studies of dampings to gravitational waves and scalar perturbations in the context of cosmological first order phase transitions
can thus help reveal equally important information of the underlying particle interactions.

\sectionname{Conclusion}
In this work, we have studied the effect of dissipations on affecting the gravitational wave production from the usually
dominant sound waves. We observe dampings of gravitational wave spectrum on the high frequency tail for short 
damping length and on a broader frequency range for longer ones where the peak frequency can also be shifted to lower
frequencies. These modifications serve as a set of new observables for probing cosmic phase transitions and break the parameter 
degeneracy that plagues previous particle physics studies, thus opening up new portals to studies of the underlying particle physics models. 
This also implies that updated searches are desired at LIGO, by the PTA experiments and others.

\sectionname{Acknowledgements}
We would like to thank Ligong Bian, Wei Chao, Yvonne Wong for helpful discussions. This work is supported by the 
startup fund provided by the University of Chinese Academy of Sciences and by the National Science Foundation of 
China (NSFC) under Grant No. 12147103..
 

\bibliographystyle{utphys}

\end{document}